\documentclass[aps,prm,preprint,amsmath,amssymb]{revtex4-2}

\usepackage{braket}
\usepackage[version=3]{mhchem}
\usepackage{graphicx}
\usepackage{upgreek} 
\usepackage{bm}

\begin{document}
\author{Vivek~Jain}
\author{Salman Ahsanullah}
\author{Neema Rafizadeh}
\author{Gbenga S. Agunbiade}
\author{Hui~Zhao}
\email{huizhao@ku.edu}
\affiliation{Department of Physics and Astronomy, The University of Kansas, Lawrence, Kansas 66045, United States}

\title{Density-dependent photocarrier recombination in $\gamma$-InSe}

%\keywords{}

%\begin{tocentry}
% \includegraphics[width=8.25cm]{toc.pdf}
%\end{tocentry}

\begin{abstract}
We studied the photocarrier dynamics in $\gamma$-InSe using femtosecond transient reflectance spectroscopy. By measuring the dependence of the differential reflectance on the injected carrier density, we established an empirical relation between the optical response and carrier density, allowing quantitative analysis of the recombination dynamics. The carrier decay exhibits a strong density dependence, indicating a substantial contribution from nonlinear recombination. At high carrier densities, the dynamics are dominated by a second-order recombination process, with a rate constant of $3.8\times10^{-6}~\mathrm{cm^3\,s^{-1}}$. At low densities, the decay becomes approximately exponential, with a carrier lifetime of about 7 ps. We also resolved an initial rise of the band-edge response with a time constant of less than 1 ps, which is attributed to hot-carrier cooling. These results reveal the competition between first-order and second-order recombination in $\gamma$-InSe and provide quantitative insights into its ultrafast photocarrier dynamics.
\end{abstract}
\maketitle

\section{Introduction}
As a van der Waals layered semiconductor, InSe has drawn considerable attention for potential applications in electronics and optoelectronics. Its crystal structure consists of covalently bonded Se–In–In–Se layers held together by van der Waals interactions. Different layer-stacking sequences give rise to several polytypes, including $\beta$-, $\epsilon$-, and $\gamma$-InSe \cite{b114185406}. Among these, $\gamma$-InSe has a rhombohedral structure. Transport studies on encapsulated $\gamma$-InSe flakes have demonstrated electron mobilities exceeding 10$^3$ cm$^2$ V$^{-1}$ s$^{-1}$ at room temperature and even higher values at liquid-helium temperatures \cite{nn12223}. The high mobility at low temperature enabled observation of the quantum Hall effect \cite{nn12223}. Its direct bandgap near 1.25 eV in the bulk form makes it attractive for near-infrared optoelectronics \cite{nn12223, afm352413672}. Optical measurements also showed that reducing the thickness substantially increases the band-edge transition energy \cite{nn12223}. These properties have stimulated exploration of this material for optoelectronic applications. Photodetectors based on multilayer $\gamma$-InSe and its graphene heterostructures have demonstrated sensitive optical detection \cite{acsami1421383, nl247252}. For light-emitting applications, bulk-like $\gamma$-InSe flakes show strong room-temperature near-infrared photoluminescence \cite{afm352413672,nl234058, nl223840}. Furthermore, the noncentrosymmetric structure of $\gamma$-InSe enables second-order nonlinear optical applications, such as second-harmonic generation \cite{nanoscale1510125}. 

Understanding photocarrier dynamics is essential for developing optoelectronic applications. Energy relaxation determines how rapidly photoexcited carriers reach the band edges, while recombination controls the carrier population available for charge collection and light emission. Time-resolved studies have investigated these processes in InSe crystals, thin flakes, and nanosheets using two-photon photoemission~\cite{b97241201}, optical pump--probe spectroscopy~\cite{jpcl10493,nt31095713,aom92101432,oe3017967,nanoscale153169}, terahertz spectroscopy~\cite{jpcl74191,b102014314}, and time-resolved photoluminescence~\cite{prm4044001}. Specifically for $\gamma$-InSe, transient reflectance measurements have revealed photoinduced bandgap renormalization and demonstrated that carrier recombination depends strongly on flake thickness and surface condition~\cite{nanoscale153169}. Polarization-resolved transient measurements have further identified anisotropic exciton screening under photoexcitation~\cite{n2dma937}. 

The competition between recombination channels is particularly important as the injected carrier density increases. Here, we investigate density-dependent photocarrier recombination in an hBN-protected, 87-nm-thick $\gamma$-InSe flake using time-resolved pump–probe reflectance spectroscopy at room temperature. We observed carrier dynamics consistent with an effective second-order recombination contribution at high injection densities. At low injection densities, the decay approaches a density-independent lifetime of approximately 7 ps, while the initial signal rise is consistent with subpicosecond carrier cooling. These results characterize the density-dependent ultrafast photocarrier dynamics in bulk-like $\gamma$-InSe.

\section{Sample Fabrication and Characterization}

\begin{figure}[t]
    \centering\includegraphics[width=16cm]{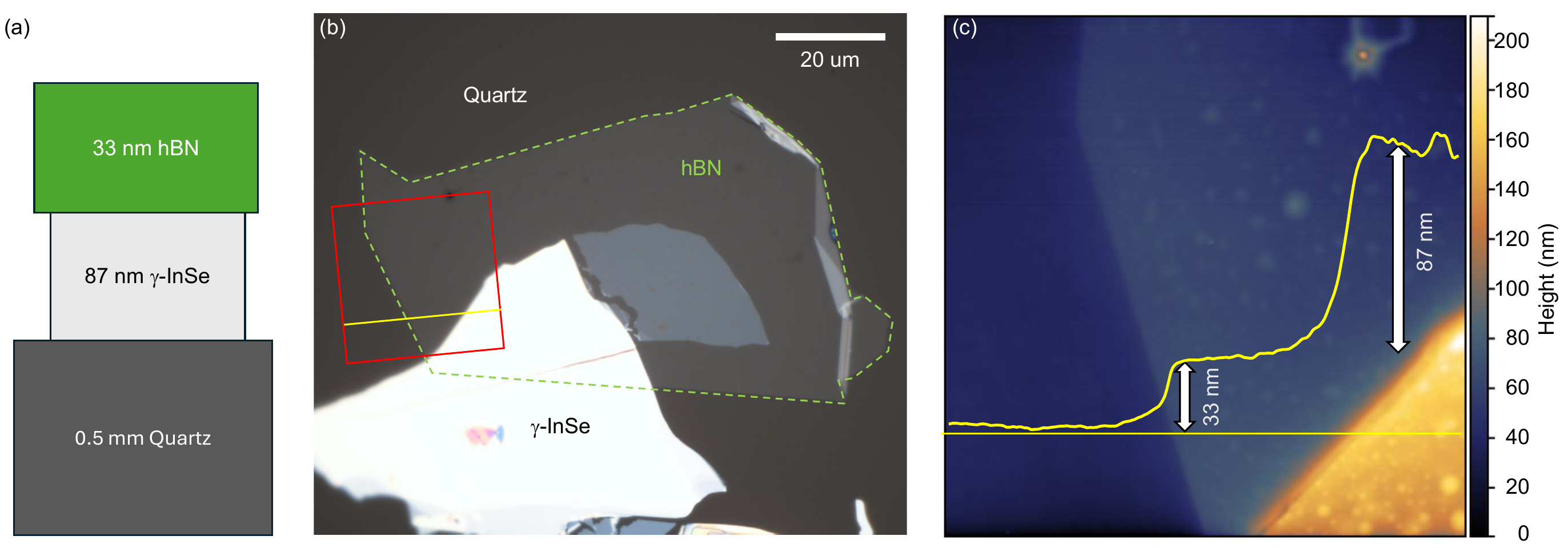}
\caption{(a) Schematic illustration of the sample structure, consisting of an 87-nm-thick $\gamma$-InSe flake on a quartz substrate, capped by a 33-nm-thick hexagonal boron nitride (hBN) flake. (b) Optical microscopy image of the sample. The boundary of the hBN flake is indicated by the green dashed line. (c) Atomic force microscopy image of the region indicated by the red box in (b). The yellow curve shows the height profile along the yellow line in (c).} 
\label{fig:samplePL}
\end{figure}

The $\gamma$-InSe sample was prepared by mechanical exfoliation. A $\gamma$-InSe crystal (purchased from 2D Semiconductors) was first exfoliated using adhesive tape to expose a fresh surface. Flakes were subsequently exfoliated from this surface onto a polydimethylsiloxane (PDMS) substrate. A large, uniform $\gamma$-InSe flake was selected and transferred onto a quartz substrate. An hBN flake was prepared using the same procedure and transferred onto the $\gamma$-InSe flake to protect it from environmental degradation. All exfoliation and transfer steps were performed in an Ar-filled glovebox. The sample was then thermally annealed to improve the interfacial contact between the flakes. Figure~\ref{fig:samplePL}(a) illustrates the sample structure, and Fig.~\ref{fig:samplePL}(b) shows an optical microscopy image of the sample.

The thicknesses of the $\gamma$-InSe and hBN flakes were determined by atomic force microscopy (AFM). Figure~\ref{fig:samplePL}(c) shows an AFM image of the region indicated by the red box in Fig.~\ref{fig:samplePL}(b), together with a height profile along the yellow line. From left to right, the profile crosses the bare quartz substrate, the portion of the hBN flake resting directly on quartz, and the region where hBN covers $\gamma$-InSe. The first height step of 33~nm corresponds to the hBN thickness, while the additional step of 87~nm gives the thickness of the underlying $\gamma$-InSe flake.

The use of an Ar-filled glovebox for the entire sample fabrication process was essential for preserving the quality and stability of the $\gamma$-InSe sample. The hBN capping layer provided additional protection against environmental degradation during optical measurements in air. The effectiveness of these precautions was confirmed by the long-term stability of the sample, which showed no noticeable signs of degradation after several months of exposure to air and repeated laser measurements. In contrast, $\gamma$-InSe flakes prepared in air or without hBN encapsulation exhibited visible degradation within a few days.

\begin{figure}[t]
    \centering\includegraphics[width=9cm]{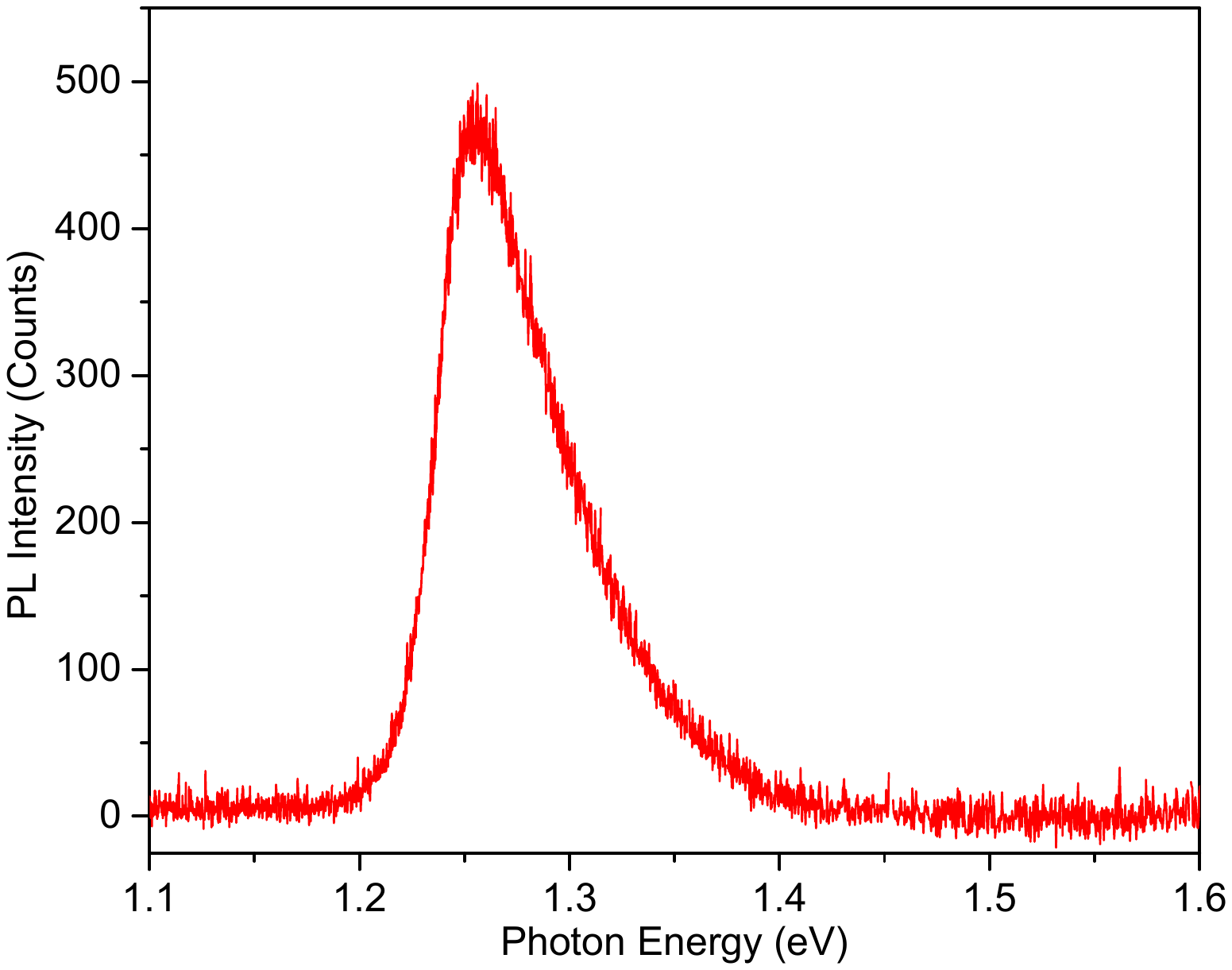}
\caption{Photoluminescence (PL) spectrum of $\gamma$-InSe measured with a 1.959-eV (632.8-nm) continuous-wave laser, after subtraction of the quartz substrate background. The emission peak is located at approximately 1.265 eV (980 nm).} 
\label{fig:PL}
\end{figure}

We performed photoluminescence (PL) measurements at room temperature. The sample was excited by a continuous-wave helium-neon laser at 632.8 nm with an incident power of 55 $\upmu$W. As shown in Fig.~\ref{fig:PL}, the PL spectrum of $\gamma$-InSe exhibits a peak at 1.265 eV (980 nm), close to the room-temperature band-edge PL peaks reported for bulk $\gamma$-InSe samples, which range from approximately 1.23 to 1.26 eV.~\cite{nn12223,acsnano161477,nl234058,afm352413672,nl259526,afm352508213,nl2613385} 

\section{Photocarrier Dynamics}

Photocarrier dynamics were studied by transient reflectance spectroscopy. Our laser system consisted of an 80 MHz Ti:sapphire oscillator (MKS Mai Tai) and an optical parametric oscillator (MKS Inspire). The 820 nm output pulses from the Mai Tai were directed into the Inspire, where they were converted to 410 nm by second-harmonic generation in a beta-barium borate (BBO) crystal to pump the optical parametric oscillator, producing tunable visible signal and near-infrared idler outputs. A small portion of the 410 nm (3.024 eV) beam was used as the pump, while the tunable idler output was used as the probe. The pump and probe beams were focused onto the sample at normal incidence using an objective, with spot sizes of approximately $6~\upmu$m. An imaging system was used to monitor the sample surface and the pump and probe spots, allowing their spatial overlap on the selected sample region. The pump-induced differential reflectance, $\Delta R/R_0=(R-R_0)/R_0$, was measured using a photodiode and a lock-in amplifier, where $R$ and $R_0$ are the sample reflectances with and without pump excitation, respectively. The pump beam was modulated at approximately 3 kHz by an optical chopper for lock-in detection~\cite{afm271604509}. All the measurements were performed with the sample in air at room temperature.

\begin{figure}[t]
    \centering\includegraphics[width=9cm]{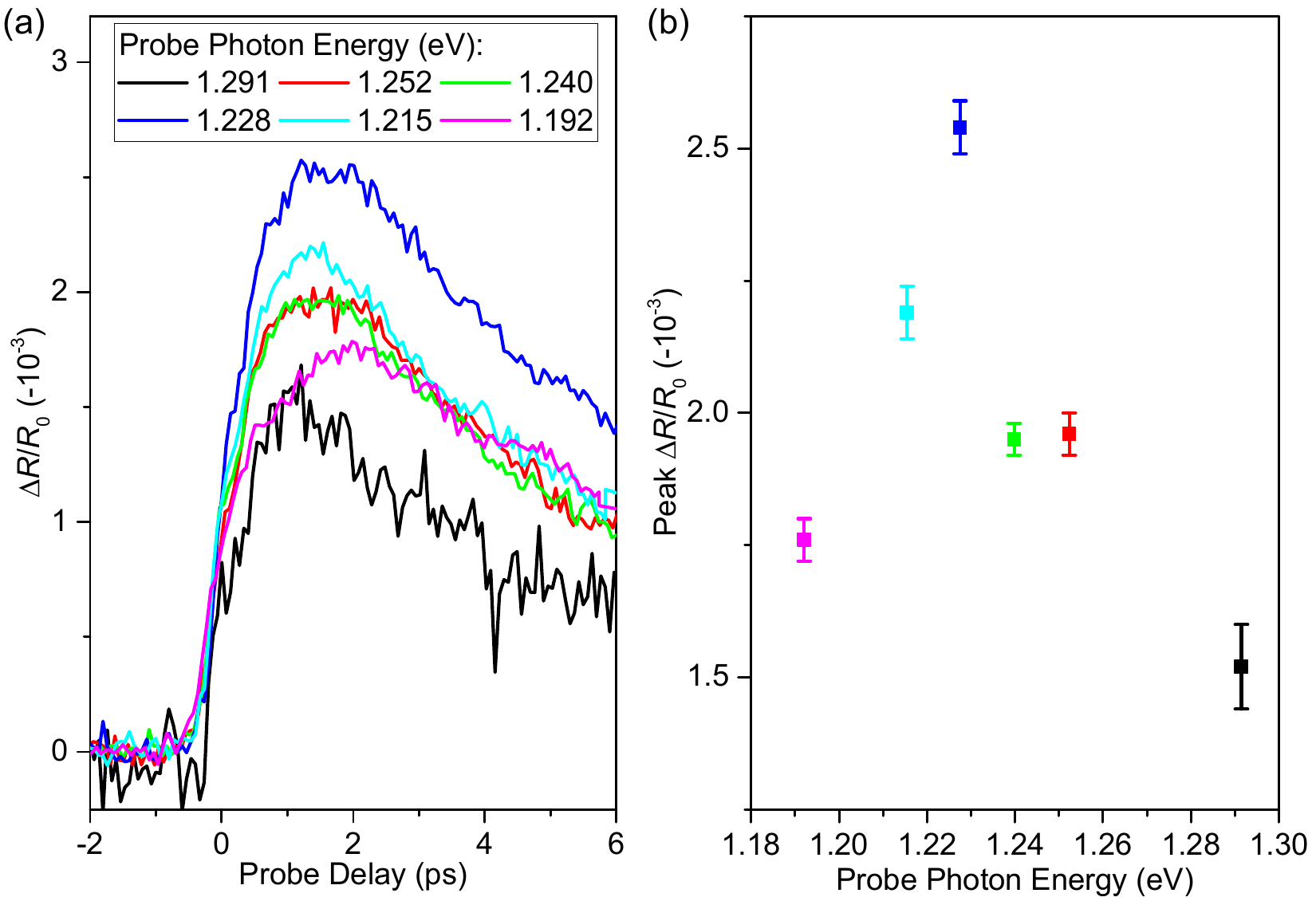}
\caption{(a) Differential reflectance of $\gamma$-InSe as a function of pump--probe delay with different probe photon energies as labeled. The negative differential reflectance is plotted with the sign reversed for clarity. The pump photon energy was fixed at 3.02 eV. (b) Peak magnitude of the differential reflectance as a function of probe photon energy. The maximum occurs at 1.228 eV (1010 nm). }
\label{fig:probeenergy}
\end{figure}

\begin{figure*}[t]
    \centering
    \includegraphics[width=16cm]{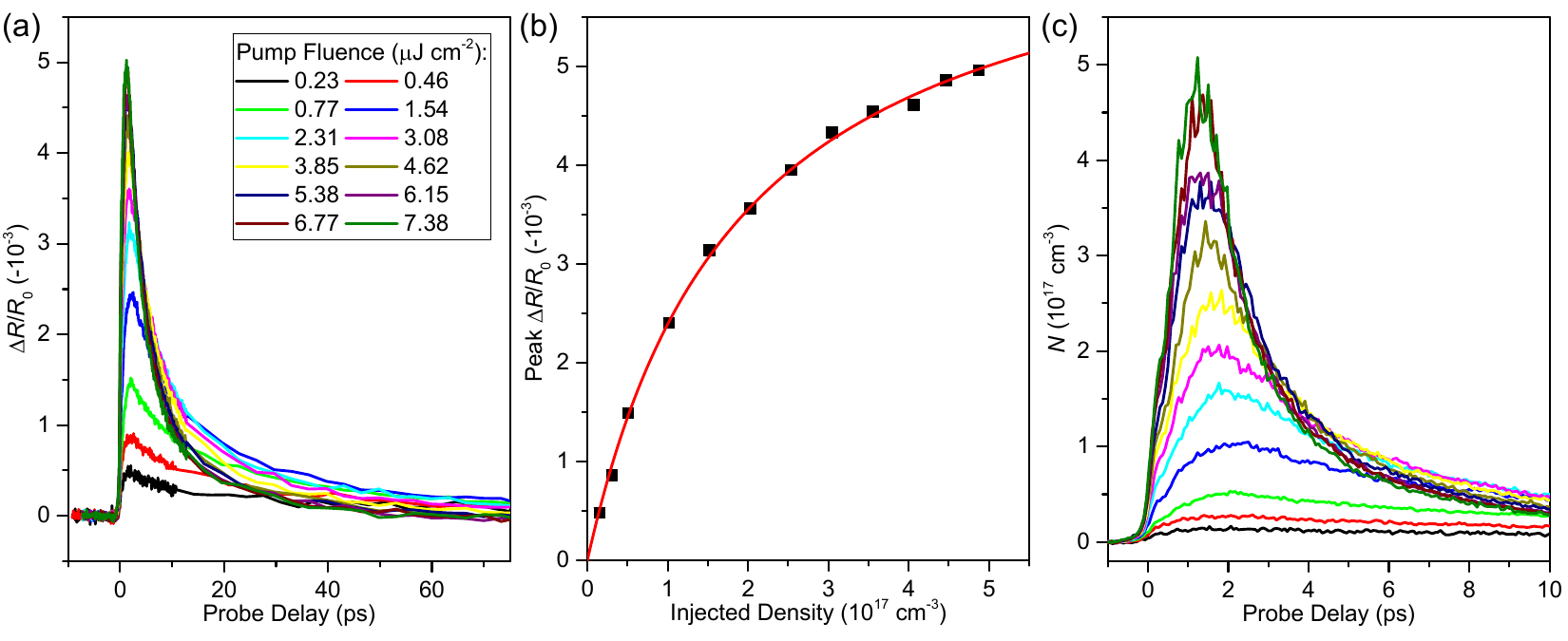}
    \caption{(a) Pump-induced differential reflectance of $\gamma$-InSe measured as a function of pump--probe delay at different pump fluences, as labeled. The pump and
    probe photon energies were 3.024 and 1.228 eV, respectively. (b) Peak magnitude of the differential reflectance as a function of the initial photocarrier density. The red curve
    is a fit using the saturation model described in the text. (c) Time evolution of the photocarrier density calculated from the differential reflectance in (a) using the
    experimentally established relation in (b).}
    \label{fig:DRoR}
\end{figure*}

We first performed differential reflectance measurements with different probe photon energies to identify the optimal probe condition. The sample was excited by a 3.024~eV pump pulse with an energy fluence of 1.5~$\upmu$J~cm$^{-2}$. Previously reported exciton binding energies in bulk-like $\gamma$-InSe range from approximately 10 to 20~meV, below the thermal energy of 26~meV at room temperature~\cite{nanoscale153169,nl234058}. Under these conditions, the photocarrier population is expected to consist predominantly of free electrons and holes, consistent with previous interpretations of room-temperature transient reflectance and photoluminescence measurements~\cite{nanoscale153169,afm352413672}. The probe photon energy was tuned around the PL peak, and the differential reflectance was recorded as a function of pump--probe delay at each energy. The results are shown in Fig.~\ref{fig:probeenergy}(a). The differential reflectance is negative at all probe energies, indicating that the pump-injected carriers reduce the reflectance of the sample. The signal rises rapidly after pump excitation, reaches a maximum at approximately 1--2 ps, and then decays. The peak magnitude of the differential reflectance is plotted as a function of probe photon energy in Fig.~\ref{fig:probeenergy}(b). The largest signal was obtained at a probe photon energy of 1.228 eV (1010 nm).

We next measured the differential reflectance using a probe photon energy of 1.228~eV and a pump photon energy of 3.024~eV at various pump fluences. The results are shown in Fig.~\ref{fig:DRoR}(a). The signal rises rapidly to a peak and then decays on a timescale of approximately 10 ps. Although the signal decay reflects the decay of the photocarrier population, an accurate description of the dynamics requires an experimentally established relation between the differential reflectance and the photocarrier density. We estimated the injected photocarrier density from the pump fluence. At the pump photon energy of 3.024 eV, the measured transmittance and reflectance of the unpumped sample were $T_0 = 0.32$ and $R_0 = 0.40$, respectively, giving an absorptance of $A_0 = 1 - T_0 - R_0 = 0.28$. Assuming that each absorbed photon generates one electron--hole pair and that the carrier density decreases exponentially with depth, the injected density at the front surface of the $\gamma$-InSe flake was estimated to be $8.9\times10^{16}~\mathrm{cm}^{-3}$ for an incident pump fluence of $1~\upmu\mathrm{J}~\mathrm{cm}^{-2}$. The corresponding density at the back surface was $4.8\times10^{16}~\mathrm{cm}^{-3}$, while the thickness-averaged density was $6.6\times10^{16}~\mathrm{cm}^{-3}$. These densities scale linearly with the incident pump fluence, as nonlinear absorption is expected to be negligible at the fluence used in this experiment.

Figure~\ref{fig:DRoR}(b) shows the peak magnitude of the differential reflectance as a function of the injected photocarrier density, calculated using the thickness-averaged density as described above. The signal increases with carrier density but exhibits clear saturation behavior, indicating that the differential reflectance is not proportional to the carrier density, $N$. We describe this dependence using a saturation model, $\Delta R/{R_0} = -S_{\mathrm{sat}}N/(N+N_{\mathrm{sat}})$, where $S_{\mathrm{sat}}$ is the saturation amplitude and $N_{\mathrm{sat}}$ is the saturation density. As shown by the red curve in Fig.~\ref{fig:DRoR}(b), the model fits the experimental data well, with $N_{\mathrm{sat}}=1.88\times10^{17}~\mathrm{cm}^{-3}$ and $S_{\mathrm{sat}}=6.89\times10^{-3}$.  This experimentally established relation allows us to convert the measured differential reflectance into the photocarrier density, as shown in Fig.~\ref{fig:DRoR}(c).

\begin{figure}[t]
    \centering\includegraphics[width=9cm]{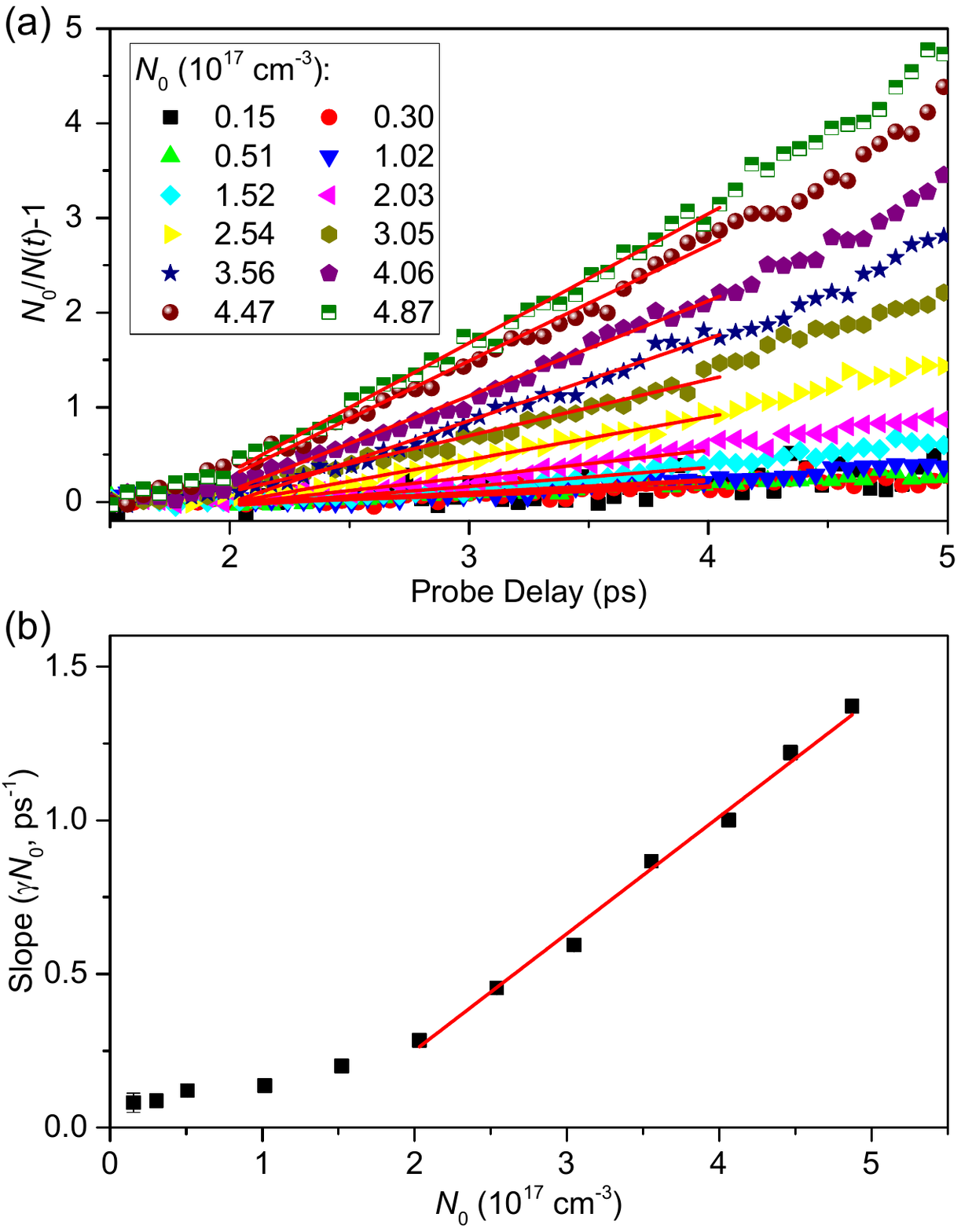}
\caption{(a) $N_0/N(t)-1$ as a function of pump--probe delay at different initial photocarrier densities as labeled, calculated from the density evolution shown in Fig.~\ref{fig:DRoR}(c). The red lines are linear fits to the early-time data. (b) Slopes obtained from the linear fits in (a) as a function of the initial photocarrier density. The red line is a linear fit to the high-density data, yielding a second-order recombination rate constant of $\gamma=3.8\times10^{-6}~\mathrm{cm^3\,s^{-1}}$.}
\label{fig:xxa}
\end{figure}

The density evolution shown in Fig.~\ref{fig:DRoR}(c) exhibits a clear density dependence of the photocarrier decay. The faster decay at higher densities indicates that the dynamics cannot be described by a linear rate equation with a constant recombination rate. Instead, nonlinear recombination processes involving multiple photocarriers must contribute to the decay. One possible mechanism is electron--hole pair annihilation, in which two electron--hole pairs interact and one pair recombines. Including this second-order recombination process, the time evolution of the photocarrier density $N(t)$ can be modeled by the rate equation
\begin{equation}
    \frac{dN}{dt}=-\frac{N}{\tau}-\gamma N^2,
\end{equation}
where $\tau$ is the carrier lifetime associated with first-order recombination and $\gamma$ is the second-order recombination rate constant. In the high-density regime, where the second-order term dominates, the first-order term can be neglected. The solution is then
\begin{equation}
    \frac{N_0}{N(t)}-1=\gamma N_0t,
\end{equation}
where $N_0$ is the initial carrier density~\cite{b89125427}. To examine this relation, we calculated $N_0/N(t)-1$ from the density evolution curves in Fig.~\ref{fig:DRoR}(c). The results are shown in Fig.~\ref{fig:xxa}(a). At early probe delays after the initial signal rise, this quantity increases approximately linearly with time. Linear fits, shown by the red lines, were used to extract the slopes, which are plotted as a function of the initial carrier density in Fig.~\ref{fig:xxa}(b). At high injection densities, the slope increases linearly with the initial photocarrier density, consistent with the second-order recombination model. A linear fit to the high-density data (red line) yields a second-order recombination rate constant of $\gamma = 3.8 \times 10^{-6}~\mathrm{cm^3\,s^{-1}}$. This value is approximately three orders of magnitude larger than the bimolecular radiative coefficient previously used to analyze the steady-state PL quantum yield of $\gamma$-InSe~\cite{afm352413672}. If this estimate applies to our flake, bimolecular radiative recombination accounts for only about $0.1\%$ of the observed second-order photocarrier decay rate.

\begin{figure}[t]
    \centering\includegraphics[width=9cm]{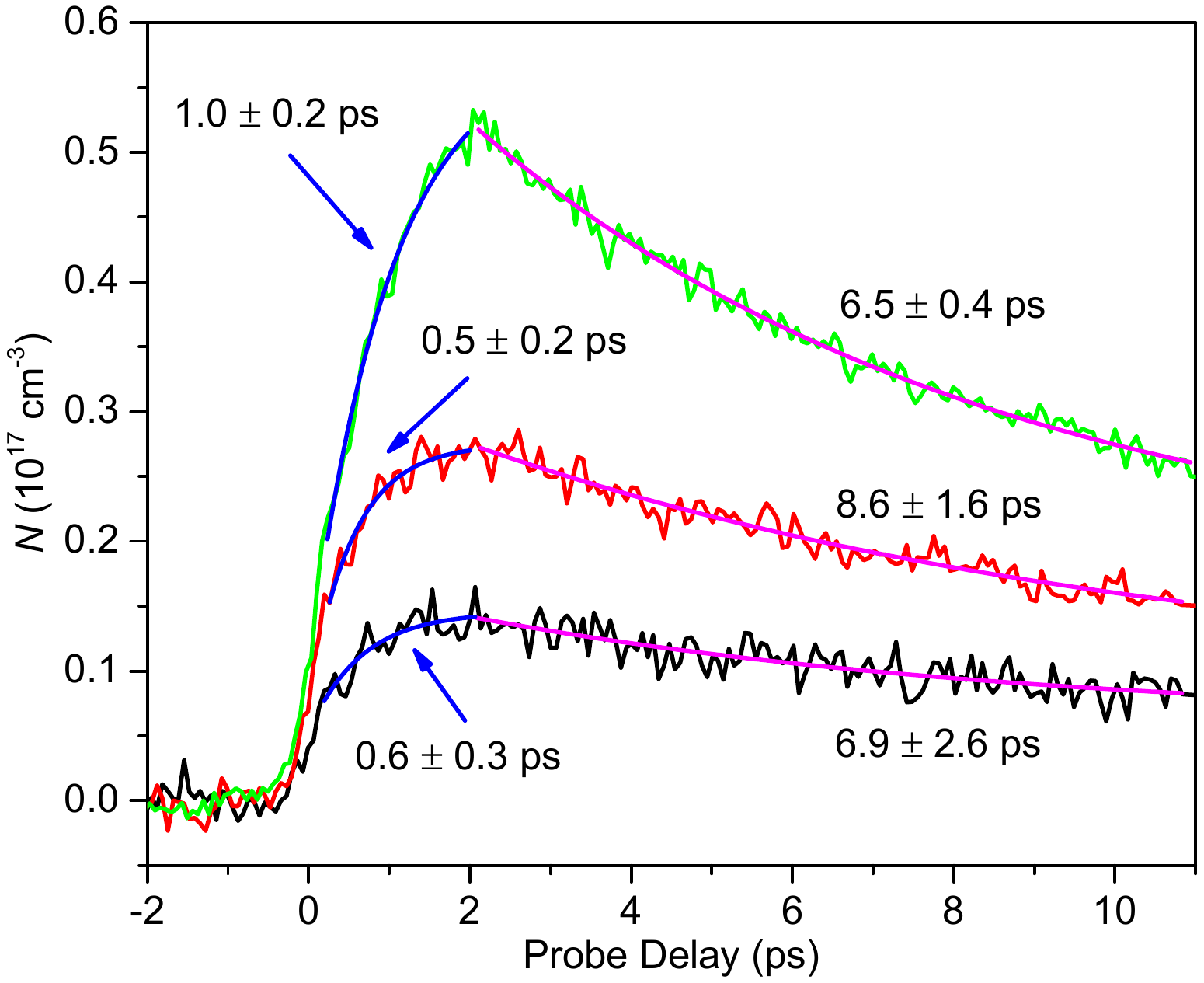}
\caption{Time evolution of the photocarrier density in $\gamma$-InSe measured at three low injection densities. The rising portions were fitted with exponential growth functions to extract the carrier cooling times, while the decaying portions were fitted with single-exponential functions to determine the recombination lifetimes. The fitted curves are shown by the solid lines, with the corresponding time constants labeled.}
\label{fig:lowdensity}
\end{figure}

The data from the low injection density measurements shown in Fig.~\ref{fig:xxa}(b) deviate from the linear dependence observed at high densities. This is expected because, at sufficiently low densities, the first-order recombination term in Eq.~(1) can no longer be neglected. However, at such low densities, the second-order recombination term may become negligible, allowing the carrier lifetime to be directly determined by exponential fits. To examine this, Fig.~\ref{fig:lowdensity} shows the carrier density evolution at three low injection densities, replotted from Fig.~\ref{fig:DRoR}(c). Single-exponential fits, $N(t)=N_0\exp(-t/\tau)$, yield decay times of 6.5–8.6 ps, with no systematic dependence on fluence within the uncertainties. We therefore obtain an effective low-injection decay time of approximately $\tau = 7$ ps. 

The initial rise of these curves is also clearly resolved. Since the carriers are excited to high-energy states by the 3.024-eV pump pulse and probed near the band edge at 1.228 eV, this rise can be attributed to the energy relaxation of hot carriers toward the band-edge states. By fitting the rising portions with exponential growth functions, we obtained time constants for this cooling process of less than 1 ps.

\section{Conclusions}

In summary, we have investigated the density-dependent photocarrier dynamics in bulk-like $\gamma$-InSe using femtosecond transient reflectance spectroscopy. By establishing an experimental relation between the differential reflectance and the photocarrier density, we were able to quantitatively analyze the recombination dynamics over a broad range of injection densities. We found that photocarrier recombination changes substantially with carrier density. At low injection densities, the carrier population exhibits approximately exponential decay with a characteristic lifetime of about 7~ps. At higher densities, nonlinear recombination becomes increasingly important, and the dynamics are consistent with an effective second-order recombination process with a rate constant of $3.8\times10^{-6}$~cm$^3$~s$^{-1}$. This large coefficient indicates that interactions involving multiple photocarriers provide an efficient carrier-loss channel under high excitation conditions. We also resolved a subpicosecond rise of the band-edge optical response, consistent with rapid cooling of photoexcited carriers.

These findings have important implications for understanding the optical and electronic properties of $\gamma$-InSe. The strong density dependence demonstrates that the photocarrier lifetime cannot be represented by a single material parameter independent of excitation conditions. Instead, the relative contributions of first-order and nonlinear recombination must be considered when evaluating carrier relaxation and recombination. In particular, the rapid nonlinear carrier loss at high injection densities may reduce the carrier population available for charge collection or light emission, potentially affecting the performance of $\gamma$-InSe-based optoelectronic devices under high injection levels.

\begin{acknowledgments}
This work was supported by the U.S. Department of Energy, Office of Basic Energy Sciences, Division of Materials Sciences and Engineering, under Award No.~DE-SC0020995.
\end{acknowledgments}

%\bibliography{References}


\end{document}